\documentclass[aps, prl, twocolumn,showpacs]{revtex4}
\usepackage{graphicx}
\newcommand\beq{\begin{equation}}
\newcommand\eeq{\end{equation}}
\newcommand\bear{\begin{eqnarray}}
\newcommand\eear{\end{eqnarray}}
\begin{document}
\title{The Nature of Vibrations of a 2-D Disordered Lattice Model with Nano-scale Order}
\author{Prasenjit Ghosh and Umesh V Waghmare}
\affiliation{Theoretical Sciences Unit,
Jawaharlal Nehru Centre for Advanced Scientific Research,\\ Jakkur PO,
Bangalore 560 064, India}
\begin{abstract}
Motivated by the unusual soft mode anomalies in the relaxor ferroelectrics, we present 
a numerical study of vibrational excitations of a disordered lattice model with chemical 
order on the nano-scale.
We find an
Ioffe-Regel crossover in acoustic modes separating propagating low-frequency modes
from the intermediate frequency diffusons, and localized modes at the high-frequency end
of the spectrum. At a fixed degree of disorder, we find increasing uncertainty in the 
momentum of long wave-length optical modes with decrease in the size of nano-ordered 
regions. We determine the cause for this to be a strong mixing between acoustic and optical
modes, believed to be central to the model of waterfall phonon anomalies.

\end{abstract}
\pacs{63.22.+m 63.20.Dj 74.25.Kc}
\maketitle

Nano-scale structure of materials has interesting effects on their 
macroscopic behavior and become yet another control parameter in
tuning material properties\cite{NanoMRS}. For example, relaxor ferroelectrics, such as
PbMg$_{\frac{1}{3}}$Nb$_{\frac{2}{3}}$O$_3$ (PMN)\cite{Cross} and
PbSc$_{\frac{1}{2}}$Nb$_{\frac{1}{2}}$O$_3$ (PSN)\cite{Setter1},
have tremendous technological potential in applications based on solid 
state transducers\cite{Service}, due to their giant electromechanical and 
dielectric response. Randomness and nano-scale regions in the ordering of 
polar fields are known to be central to these interesting 
properties\cite{Blinc, Westphal}.  In contrast to conventional ferroelectrics 
such as PbTiO$_3$ with a sharp ferroelectric transition, relaxors exhibit a 
diffuse ferroelectric transition, with a broad peak in dielectric constant 
as a function of temperature.

Ferroelectricity in conventional ferroelectric materials is essentially
a phonon-related property characterized by a soft mode near the ferroelectric
phase transition\cite{Cochran, Anderson}. The soft phonon anomalies in
the context of relaxors\cite{Gehring} were observed as 
``waterfall phonons" recently using neutron scattering, with a speculation 
that these are linked with nano-polar regions. Further studies of the 
``waterfall" phenomena based on neutron scattering were explained
using a mode-coupling theory\cite{Gehring1,Hlinka} for the coupling between 
acoustic and optical modes. While the work of Hlinka et al \cite{Hlinka} 
argued that the wave-vector $k$ of the waterfall phonon is unlikely to be related
to the size of nano-polar regions, the origin of coupling between acoustic
and optical modes is not known.

A relaxor, for example PSN, PST etc., is a solid solution with a degree of disorder 
arising from the chemical ordering of Sc and Nb (for PSN) and Sc and Ta (for PST) 
at the $B$ sites of ABO$_3$ perovskite lattice. Correlation between such disorder 
and relaxor properties of PSN and PST were shown experimentally\cite{Setter1, Setter2}.
A region of the size of $\approx~$2-6 nm\cite{Krause, Perrin, Jin} with a specific 
kind of chemical ordering defines a nano-ordered region (NOR) with intense local polar 
fields\cite{Westphal, Quian}.  Through the coupling with ferroelectric dipoles, 
they give rise to nano-polar regions (NPR), speculated to be linked with phonon 
anomalies called ``waterfall phonons" in relaxors. Precise knowledge of the nature 
of phonons in the presence of chemically nano-ordered regions is essential to establish 
their possible link with the ``waterfall phonons".

In this letter, we determine the nature of vibrational excitations in a model system
which consists of chemically-ordered nano-regions in a percolating disordered matrix.
While the model used in our analysis is a great simplification of relaxors, it allows us
to determine some general aspects of phonons in a nano-ordered system. In particular,
we find localization of modes at high frequencies and an Ioffe-Regel crossover at low 
frequencies, similar to that found in amorphous systems\cite{Ioffe, Allen, Taraskin}. 
Our findings about the nature of long-wavelength optical modes helps in uncovering 
possible links between nano-scale order and waterfall phonons in terms of the mode 
coupling analysis of Hlinka \textit{et al}\cite{Hlinka}.

For the ordered case, our model system consists of two types of atoms A and B occupying 
(0,0)$a$ and ($\frac{1}{2},\frac{1}{2}$)$a$ sites of a square lattice respectively, 
$a$ being the lattice constant. 
The interaction between two atoms displaced from their mean position by 
$\mathbf{u_i}$ and $\mathbf{u_j}$ is given by 
\beq
U_{ij}=k_L(\mathbf{\hat{b}_{||}}.\mathbf{u_i})(\mathbf{\hat{b}_{||}}.
\mathbf{u_j})+k_T(\mathbf{\hat{b}_{\perp}}.\mathbf{u_i})
(\mathbf{\hat{b}_{\perp}}.\bf{u_j})
\eeq 
where $\mathbf{\hat{b}_{||}}$ and $\mathbf{\hat{b}_{\perp}}$ 
are the unit vectors parallel  (longitudinal) and perpendicular (transverse) to
the line joining two atoms respectively, $\it{k_L}$ and $\it{k_T}$ being the 
longitudinal and transverse force constants respectively. 
Only harmonic interactions up to the second nearest neighbors ((1,1)$a$) have  
been included, giving 10 distinct interatomic force constants $k^{AAnn}_{L,T}$, 
$k^{AA2nn}_{L,T}$, $k^{BBnn}_{L,T}$, $k^{BB2nn}_{L,T}$, $k^{ABnn}_{L,T}$ 
 and two onsite interactions $k_A$ and $k_B$, 
where $nn$ and $2nn$ stand for the nearest and next nearest neighbors 
respectively. Values of the force constants (Table I) have been chosen such 
that the acoustic sum rule is obeyed, there are stable modes throughout the 
Brillouin zone, and acoustic and optical branches are well-separated.

Disorder is introduced by randomly selecting two different types of atoms
and interchanging their positions making sure that a given atom is
interchanged only once. A configuration with nano-scale order is obtained by
excluding such an interchange in the circular nano-ordered regions (NORs)
with a fixed diameter of a few nanometers. Centres of these circular regions are taken
randomly, but avoiding any overlap between them.
The degree of disorder is quantified with a disorder 
parameter $\delta$ defined as 
\beq
\delta=\frac{\mbox{No. of A-A or B-B nearest neighbors}}
{\mbox{Total no. of nearest neighbor}}.
\eeq
For a fully ordered case, $\delta=0.0$, whereas $\delta$ is $0.5$ for the completely 
disordered case. For a disordered configuration A and B occupy one of the two
sites of the square lattice with equal probability, resulting in new types of
neighboring pairs of atoms, such as (A-B)$_{2nn}$ and (A-B)$_{3nn}$. The force 
constants for their interactions are taken as (k$_X^{AAnn}$+k$_X^{BBnn}$)/2.0 and 
(k$_X^{AA2nn}$+k$_X^{BB2nn}$)/2.0 respectively, where $X=L, T$.
Similarly, nearest neighbor atoms of the same type can possibly be separated
by (0.5,0.5)$a$, for which the force constants are given by
k$_{X,disorder}^{YYnn}=$2 k$_X^{YY2nn}$, where $Y=A, B$.  The (Y-Y)$_{nn}$ 
and (Y-Y)$_{2nn}$ interactions for the ordered case are now the (Y-Y)$_{2nn}$ and 
(Y-Y)$_{3nn}$ respectively. The acoustic sum rule is imposed by correcting the diagonal
elements of the force constant matrix\cite{Gonze}. 
Using this model, we study vibrational modes of configurations
consisting of 32 $\times$ 32 lattice sites (2048 atoms) with different number of
nano-ordered regions and $\delta\approx0.4$.

\begin{table}[h]
\caption{Force constants (in ar. units) for different interactions along 
different directions}
\begin{center}
\begin{tabular}{ccccc}
\hline
\hline
Force Constants&Values&&Force Constants&Values\\
\hline
k$_A$&19.0&&k$_L^{BBnn}$&-2.5\\
k$_B$&16.0&&k$_T^{BBnn}$&-0.5\\
k$_L^{AAnn}$&-2.0&&k$_L^{BB2nn}$&-0.25\\
k$_T^{AAnn}$&-0.5&&k$_T^{BB2nn}$&0.0\\
k$_L^{AA2nn}$&-1.25&&k$_L^{ABnn}$&-2.25\\
k$_T^{AA2nn}$&0.0&&k$_T^{ABnn}$&-1.25\\
\hline
\end{tabular}
\end{center}
\end{table}

\begin{figure}
\centering
\includegraphics[scale=0.55]{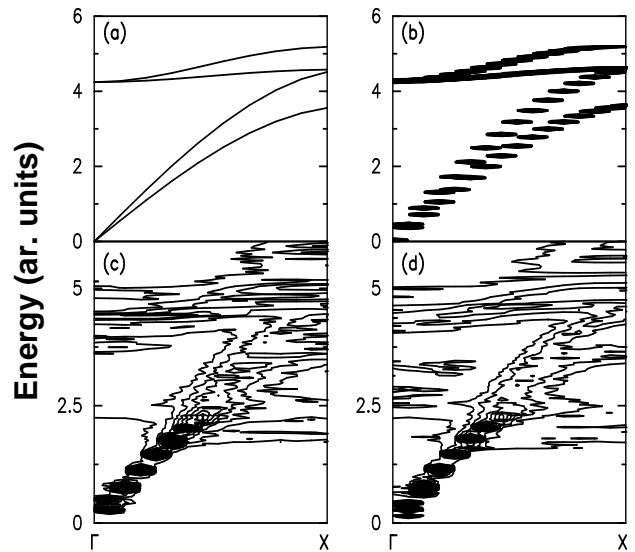}
\caption{Phonon dispersion obtained from (a) the dynamical matrix of ordered system, and
(b, c, d) contour plots of $S_{\bf{k}}(\omega)$ for the ordered system,  
a nano-ordered system with fourteen NORs each of radius $\approx$ 2.2 $a$ and a 
fully disordered system respectively.}
\label{disp}
\end{figure}

The spectral density function (S$_k(\omega)$) is calculated for different 
configurations, which allows us to relate to the constant energy scan and constant 
momentum transfer scan obtained in the neutron scattering experiments. We use the 
contour plot of S$_k(\omega)$ in visualizing the dispersion of phonon modes for systems 
with disorder. The spectral density function\cite{Schwarts} is given by
\beq
\label{sk}
S_{\mathbf{k}}(\omega)=\frac{\Sigma_{j=1}^{2N}
\vert\bar{\alpha}_{\mathbf{k}}^{j}\vert^2\delta(\omega-\omega^j)}{2N}
\eeq
 where N is the total number of atoms.
$\omega^j$ is the frequency of the $j^{th}$ eigenmode, $\vert\bar{\alpha}^j_{\bf{k}}\vert$ is
the Fourier transform of the real space eigenvector of the $j^{th}$ mode. 
Phonon dispersion of the ordered system along the $\langle10\rangle$ direction obtained from
dynamical matrix and from $S_k(\omega)$ are compared in in Fig. \ref{disp}.
At $X$ point, the longitudinal optical branch has the highest energy, whereas the
transverse optical branch is degenerate with the longitudinal acoustic branch.
Finite width of the peaks in the contour plot of $S_k(\omega)$ arise from the width of 
Gaussian used to approximate the delta function in Eqn (\ref{sk}).
Based on a reasonable agreement between the two descriptions, we use $S_k(\omega)$-based
description in the study of phonons of systems with disorder.
%
Our results for phonons of nano-ordered and disordered configurations 
(Fig. \ref{disp}(c)-(d)) show that the introduction of disorder
results in mixing of the phonon modes and subsequent broadening of peaks
in S$_k(\omega)$ as a function of both $k$ and $\omega$, $k$ is no longer a good 
quantum number to describe these modes. 
While new modes with intermediate energies appear near the $\Gamma$ point, 
long wavelength acoustic modes for $k \lesssim 0.22$ are least affected. This 
is consistent with the error-bars on phonon measurements using inelastic neutron 
scattering technique\cite{Gehring}.

\begin{figure}
\centering
\includegraphics[scale=0.35,angle=-90]{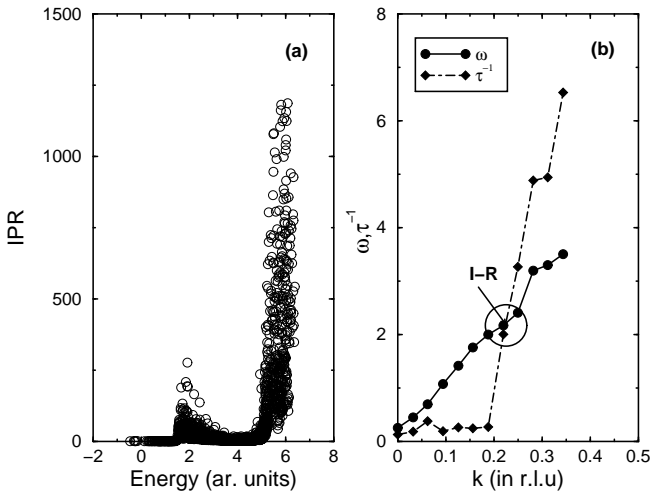}
\caption{For the configuration with 14 NORs: (a) IPR as a function of energy and
(b) acoustic mode frequency and its inverse decay time as a function of wavevector,
point of their crossover is indicated by a circle.}
\label{IR}
\end{figure}

To probe extended versus localized nature of modes in the presence of disorder and
nano-ordered regions we calculate inverse participation ratio (IPR): 
\beq
IPR(\omega_j)=N \frac{\Sigma_{\imath}\vert\vec{\alpha}_i^j\vert^4}
{(\Sigma_{\imath}\vert\vec{\alpha}_i^j\vert)^2},
\eeq
where $\vec\alpha^j$ are the eigenstates of the phonon modes. IPR is $N$ ($1$) for a fully 
localized (extended) mode. For a configuration with fourteen nano-ordered regions,
IPR as a function of energy (Fig. \ref{IR}(a)) is found to be quite similar to that 
obtained for amorphous Si\cite{Allen}. The high energy modes, between 5.0 and 6.0 units of energy, 
are highly localized and are called locons, while the rest are 
called extendons\cite{Allen}. Some modes around $\epsilon=2.0$ have intermediate values of IPR
exhibiting moderate localization. We investigated these modes further by examining distribution
of number of atoms as a function of their contribution to the IPR and find that they are
similar to the localized modes with the same IPR and $5 < \epsilon < 6$. However, the former
have non-negligible character of long-wavelength modes (evident in Fig 1) and the latter 
do not. 

Extendons can be classified into two categories, the propagons (these
propagate ballistically) and the diffusons. Wave vector $k$ is a good quantum 
number for propagons, but not for diffusons.
The Ioffe-Regel (I-R) cross-over \cite{Ioffe} separates propagons and diffusons
as a function of frequency.  In the frequency domain, the full width at half maxima
($\zeta$) of a peak in the spectral density function is used to obtain
the decay time $\tau$ of a propagating mode due to scattering
($\tau_k^{-1}\simeq\pi \zeta_{\omega}(k)$)\cite{Taraskin}. The crossover point between
$\omega$ vs. $k$ and $\tau^{-1}$ vs. $k$ gives the location ($k$) of
the I-R cross-over. For the configuration with fourteen nano-ordered regions (Fig. \ref{IR}(b)),
we find the I-R crossover at $k=0.22$ r.l.u. and $\omega=2.2$. 
The vibrational modes with energy between 2.2 and 5.0 are thus diffusons.

To further affirm this classification of modes, we studied their 
level spacing distribution (LSD). LSD, with its foundation in the random matrix theory (RMT), 
is known to exhibit Wigner-Dyson statistics for diffusons and Poisson statistics for locons
\cite{Allen}.
We calculated level spacings distribution using \textit{spectral unfolding},
in which the original spectrum is mapped into an ordered set of numbers with uniform density of 
states. We find that the LSD for modes with $2.2 < \epsilon < 5.0$ obeys Wigner-Dyson statistics
whereas that for modes with $\epsilon > 5.0$ obeys Poisson statistics with slight deviation
near zero spacing. This supports our classification of these modes as diffusons and locons
respectively.
  
\begin{figure}[h]
\centering
\includegraphics[scale=0.35,angle=-90]{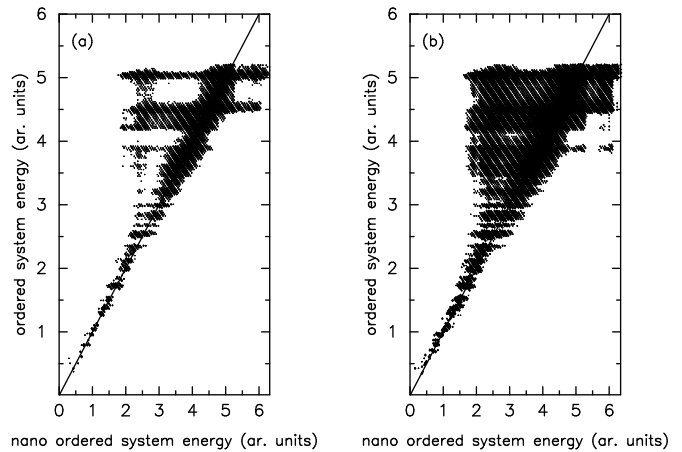}
\caption{Energy correlation between the modes of ordered system and that of 
a configuration having (a) 8 nano-ordered regions and (b) 14 nano-ordered regions. 
The straight line shows perfect correlation or {\it NO} mixing.}
\label{corr}
\end{figure}

To explore mixing among phonon modes due to disorder, we introduce 
an energy correlation function $f(\epsilon,\epsilon^{'})$ as an overlap between
the modes of the ordered system ($\vert\alpha_n^O\rangle$) 
with those of the disordered system ($\vert\alpha_m^D\rangle$):
\beq
f(\epsilon,\epsilon^{'})=\Sigma_{m,n}\vert\langle\alpha_m^D\vert\alpha_n^O
\rangle\vert^2\delta(\epsilon-\epsilon_m)\delta(\epsilon^{'}-\epsilon_n).
\eeq
It is based on the representation of modes of a disordered configuration as a linear
combination of normal modes of a 
fully ordered system. The contour plots of $f(\epsilon,\epsilon^{'})$ (Fig. \ref{corr}) 
clearly
show that the acoustic modes of a nano-ordered system in the IR-crossover region
emerge from mixing of acoustic and optical modes of the ordered system. At the
fixed degree of disorder, this mixing becomes stronger with the number of nano-ordered
regions. Secondly, diffuson modes in the optical region of the spectrum of a 
nano-ordered system arise mainly from mixing among optical modes with different wave
vector. We find that the effects of
disorder become stronger as the number (size) of nano-ordered regions increases
(decreases).

\begin{figure}
\centering
\includegraphics[scale=0.35,angle=-90]{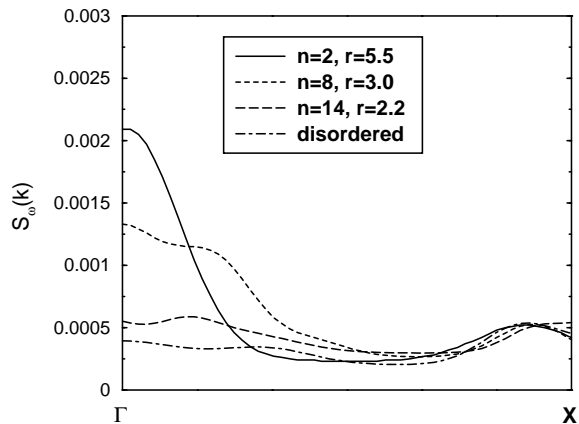}
\caption{Probability of finding an optical mode in the long wavelength limit.
``n" denotes the no. of nano-ordered regions present in the system and ``r" is the radius
of each nano-ordered region. These modes have energy $\omega$=4.292$\pm$0.034 units. }
\label{GP}
\end{figure}

To understand the nature of diffusons, we show in Fig. \ref{GP} the
$k-$dependence of $S_{\omega}(k)$ for $\omega \simeq 4.292 \pm 0.034$ corresponding
to peaks of $S$ at the $\Gamma$ point for a given configuration. It shows that 
the probability of finding a long wavelength optical mode decreases as the number 
of nano-ordered regions increases, as many of these degrees of freedom precipitate
into the acoustic branches due to strong mixing.
Correspondingly, the uncertainty in momentum $\hbar k$, which is given by the full width 
at half maximum (FWHM) of a peak in spectral-density also increases.
For configurations having fourteen or more nano-ordered regions, there is no well-defined
peak for which the FWHM can be estimated.

Due to scattering from chemical disorder, optical modes (propagons) of the ordered 
system become diffusive. We suggest that the diffusons should be relevant to
the diffuse dynamical properties of relaxors or diffuse phase transition behavior 
in ferroelectrics. While the phonons considered here are hard (high frequency)
and treated within harmonic approximation, those in real ferroelectrics are soft and exhibit
strong anharmonicity. Scattering of soft modes due to disorder is expected to be even stronger 
and their anharmonicity will lead to scattering among phonons.  As the present work 
is on a 2-D system and some of its features are likely to be qualitatively
different in 3-D, precise implications for the real relaxors need further
investigation. 

In connection with the waterfall phonon anomaly, our results show that nano-ordered
regions of large enough size {\it are} essential for the observation of phonons with
a definite wave-vector. Secondly, their size does affect the strength of mixing
between optical and acoustic modes, though not the wavevector corresponding to
the I-R crossover. The former is the basis of mode-coupling theory and latter
is consistent with the conclusion of Hlinka et al\cite{Hlinka}. We note that
the NPR's (not NORs) are present in conventional ferroelectrics too (like the chain-like dynamic
polar distortions in KNbO$_3$\cite{kno}) over a narrow temperature range above $T_c$ and
``waterfall phenomenon" should be more general. As the acoustic modes
essentially span configurations with inhomogeneous strain, the coupling of inhomogeneous
strain with polar optical modes seems to be the common origin of ``waterfall phonons"
and also the relaxor dynamics\cite{Glinchuk}.

In summary, we find that chemical disorder on a 2-D lattice results in phonon modes 
of type: propagons, diffusons and locons. At the fixed degree of disorder, reduction in
the size of nano-ordered regions results in increase (decrease) in the width (height) of
peaks corresponding to long wavelength optical modes as a function of $k$. This 
correlates with corresponding increase in the mixing between optical and acoustic
modes. While the length-scale of chemical ordering needs to be large enough (a few 
nano-metre) for observing optical phonon modes with a definite momentum, it also 
has to be small to give adequate mixing with acoustic modes responsible for the
``waterfall phonons".

We thank Chandrabhas Narayan, Ranjith and Srikanth Sastry for useful discussions. 
UVW acknowledges a DuPont young faculty award and PG acknowledges CSIR, India for 
a research scholarship.

\end{document}